\documentclass[prd,aps,amssymb,showpacs]{revtex4}

\newcommand{\beq}{\begin{equation}}
\newcommand{\eeq}{\end{equation}}
\newcommand{\bear}{\begin{eqnarray}}
\newcommand{\ear}{\end{eqnarray}}
\newcommand{\earn}{\nonumber \end{eqnarray}}

\newcommand{\nn}{\nonumber \\}
\newcommand{\Ref}[1]{(\ref{#1})}

\newcommand{\phisq}{\langle\phi^2\rangle}

\newcommand{\mds}{m_{ \mbox{\tiny \sl DS}}}
\newcommand{\gsim}{\mathop{\lefteqn{\raise.9pt\hbox{$>$}}
\raise-3.7pt\hbox{$\sim$}}}
\newcommand{\lsim}{\mathop{\lefteqn{\raise.9pt\hbox{$<$}}
\raise-3.7pt\hbox{$\sim$}}}
\arraycolsep=\mathsurround

\begin{document}

\title{Analytical approximation for ${\langle \varphi^2 \rangle}$
of a quantized scalar field in ultrastatic asymptotically flat
spacetimes}

\author{Arkady A. Popov}
\email{popov@kspu.kcn.ru}
\affiliation{Department of Mathematics,
Kazan State Pedagogical University, Mezhlauk 1 st., Kazan 420021,
Russia}

\begin{abstract}
Analytical approximations for ${\langle \varphi^2 \rangle}$ of a
quantized scalar field in ultrastatic asymptotically flat
spacetimes are obtained. The field is assumed to be both massive
and massless, with an arbitrary coupling $\xi$ to the scalar
curvature, and in a zero or nonzero temperature vacuum state.  The
expression for ${\langle \varphi^2 \rangle}$ is divided into low-
and high-frequency parts. The expansion for the high-frequency
contribution to this quantity is obtained. This expansion is
analogous to the DeWitt-Schwinger one. As an example, the
low-frequency contribution to ${\langle \varphi^2 \rangle}$ is
calculated on the background of the small perturbed flat spacetime
in a quantum state corresponding to the Minkowski vacuum at the
asymptotic. The limits of the applicability of these
approximations are discussed.

\end{abstract}


\maketitle


\section{Introduction}

The investigations of black hole evaporation and particle
production by an expanding universe have acted as a stimulus for a
detailed and systematic investigation of the theory of quantum
fields propagating on curved spacetimes. The main objects to
calculate from quantum field theory in curved spacetime are the
quantities $\left< \varphi^2 \right>$ and $\left< T^{\mu}_{\nu}
\right>$ where $\varphi$ is the quantum field and $T^{\mu}_{\nu}$
is the stress-energy tensor operator for $\varphi$. The
renormalized stress-energy tensor $\left< T^{\mu}_{\nu} \right>$
is an important quantity for the construction  of a
self-consistent model of an evaporating black hole, while the
mean-square field $\left< \varphi^2 \right>$ plays a role in the
study of theories with spontaneous symmetry breaking. The
functional dependence $\left< T^{\mu}_{\nu} \right>$ on metric
$g_{\mu \nu}$ allows us to study the evolution of the background
geometry driven by the quantum fluctuation of the matter fields
propagating on it. This is the so-called backreaction, governed by
the semiclassical Einstein equations
      \beq
      G^{\mu}_{\nu} =8 \pi \langle T^{\mu}_{\nu}
      \rangle.
      \eeq
However, the exact results for $\left< \varphi^2 \right>$ and
$\left< T^{\mu}_{\nu} \right>$ in four dimensions are not numerous
(see, for example, \cite{DC}). Numerical computations of these
quantities are as a rule extremely intensive \cite{C,AHS,GAC}.

One of the most widely used techniques to obtain information about
these quantities is the DeWitt-Schwinger (DS) expansion \cite{DW}.
It may be used to give the expansions for $\langle \varphi^2
\rangle$ and $\langle T^{\mu}_{\nu} \rangle$ in terms of powers of
the small parameter
      \beq \label{mL}
      \frac1{m L} \ll 1,
      \eeq
where $m$ is the mass of the quantized field and $L$ is the
characteristic scale of change of the background gravitational
field \cite{MatK}.

The analytical approximations to ${\langle \varphi^2 \rangle}$ and
${\langle T^{\mu}_{\nu} \rangle}$ for the conformally coupled
massless fields \cite{AHS,GAC,Page,Z,FZ,BF,PS,Popov} give good
results. Nevertheless, there still remains the problem of the
extension of this type approximations' applicability limits. If
the quantum field is massive but the mass of the field does not
satisfy the condition \Ref{mL} the analytical approximations to
${\langle \varphi^2 \rangle}$ and ${\langle T^{\mu}_{\nu}
\rangle}$ are even less numerous
\cite{AHS,GAC,PS,Popov,BFNFSZ,BFNS}.

In this paper, approximate expressions for ${\langle \varphi^2
\rangle}_{ren}$ of a quantized scalar field in ultrastatic
asymptotically flat spacetimes are derived. The field is assumed
to be both massless or massive with an arbitrary coupling $\xi$ to
the scalar curvature $R$, and in a zero or nonzero temperature
vacuum state. The expression for ${\langle \varphi^2
\rangle}_{ren}$ is divided into low- and high-frequency parts. The
Bunch-Parker approach \cite{BP} is used for derivation of
high-frequency contributions (HFC) to these quantities. As in the
case of massless field the quantum state of the field with mass $m
\ll 1/L$ is essentially determined by the topology of spacetime
and the boundary conditions. In this paper such dependence is
determined by the low-frequency contribution (LFC). As an example,
this contribution is calculated on the background of the small
perturbed flat spacetime in quantum state corresponding to the
Minkowski vacuum at the asymptotic.

The paper is organized as follows. In Sec. II the expressions for
the Euclidian Green's function of a scalar field with arbitrary
mass and curvature coupling in a ultrastatic spacetime is divided
into low- and high-frequency parts. In Sec. III the WKB
approximation of the high-frequency contribution to ${\langle
\varphi^2 \rangle}$ is derived. The low-frequency contribution is
derived and the renormalization procedure is described in Sec. IV.
The results are summarized in Sec. V. In the Appendix the
expanding of renormalization counterterm for ${\langle \varphi^2
\rangle}$ in powers of coordinate difference of the separated
points is described. The units $\hbar = c = G = k_B = 1$ are used
throughout the paper. The sign conventions are those of Misner,
Thorne and Wheeler \cite{MTW}.


\section{Green's function }


The metric in Euclidean section for an ultrastatic spacetime is
given by
  \beq\label{metric}
  ds^2= d\tau^2+g_{a b} dx^{a} dx^{b},
  \eeq
where $\tau=i t$ is the Euclidean time, $g_{a b}$ are arbitrary
functions of spatial coordinates $x^1, x^2, x^3$. (Latin indices
run from $1$ to $3$, Greek indices run from $0$ to $3$.)

In this paper, the point-splitting method is employed for the
regularization of ultraviolet divergences. When the points are
separated one can show that
        \bear \label{phi2}
        \langle \varphi^2 \rangle_{unren}&=&
        G_{\mbox{\tiny E}}(x^{\mu},\tilde x^{\nu}).
        \ear
The Euclidean Green's function $G_{\mbox{\tiny E}}$ is a solution
of the equation
        \bear
        \left(
        \Box_{x}-m^2-\xi R(x^a) \right)G_{\mbox{\tiny E}}
        (x^{\mu},\tilde x^{\nu})
        &=&-\frac{\delta(\tau, \tilde \tau)
        \delta^{(3)}(x^{a}, \tilde x^{b})}{\sqrt{g^{(3)}(x^{a})}},
        \ear
where $m$ is the mass of the scalar field, $\xi$ is its coupling
to the scalar curvature $R(x^a)$, $g^{(3)}(x^{a}) = \det g_{ab}
(x^1,x^2,x^3)$.

If the scalar field is at zero temperature then $\delta(\tau,
\tilde \tau)$ and $G_{\mbox{\tiny E}}(x^{\mu},\tilde x^{\nu})$ can
be expanded as
       \beq \label{delta0}
       \delta(\tau, \tilde \tau)=\frac1{2 \pi}
       \int_{-\infty}^{+\infty} d \omega e^{i \omega (\tau-\tilde
       \tau)},
       \eeq
       \beq \label{G0}
       G_{\mbox{\tiny E}}(x^{\mu},\tilde x^{\nu})=\frac1{2 \pi}
       \int_{-\infty}^{+\infty} d \omega e^{i \omega (\tau-\tilde
       \tau)} \tilde G_{\mbox{\tiny E}}(\omega; x^{a},\tilde
       x^{b}).
       \eeq
If the field is at temperature $T$, then the Green's function is
periodic in $\tau-\tilde \tau$ with period $1/T$. In this case
$\delta(\tau, \tilde \tau)$ and $G_{\mbox{\tiny E}}(x^{\mu},\tilde
x^{\nu})$ have the expansions
       \beq \label{delta0T}
       \delta(\tau, \tilde \tau)= T \sum^{\infty}_{n=-\infty}
       \exp[i \omega_n (\tau-\tilde\tau)],
       \eeq
       \beq \label{G0T}
       G_{\mbox{\tiny E}}(x^{\mu},\tilde x^{\nu})=T
       \sum^{\infty}_{n=-\infty} \exp [i \omega_n (\tau-\tilde\tau)]
       \tilde G_{\mbox{\tiny E}}(\omega_n; x^{a},\tilde x^{b}),
       \eeq
where $\omega_n=2 \pi T n$. In both cases  $\tilde G_{\mbox{\tiny
E}}(\omega; x^{a},\tilde x^{b})$ satisfies the equation
        \bear \label{beq}
        \frac{1}{\sqrt{g^{(3)}(x^c)}} \frac{\partial}{\partial x^a}
        \left( \sqrt{g^{(3)}(x^c)} g^{a b}(x^c) \frac{\partial}{\partial
        x^b} \tilde G_{\mbox{\tiny E}}(\omega; x^{a},\tilde x^{b}) \right)
        -\left(\omega^2+m^2+\xi R \right) \tilde G_{\mbox{\tiny
        E}}(\omega; x^{a},\tilde x^{b}) \nn
        =-\frac{\delta^{(3)}(x^{a}, \tilde x^{b})}{\sqrt{g^{(3)}(x^{a})}}.
        \ear
In the case $m \gg 1 / L$, where $L$ is a characteristic scale of
the variation of the background gravitational field, it is
possible to construct the iterative procedure of the solution of
this equation with small parameter $1/mL$ \cite{DW,BP}. This
procedure gives the standard expansion of $\left< \varphi^2
\right>_{unren}$ in terms of powers of $mL$.

In the case
        \beq \label{conm}
        m \lsim \frac{1}{L}
        \eeq
a small parameter of iterative procedure does not exist.
Nevertheless, let us divide $G_{\mbox{\tiny E}}$ into low- and
high-frequency parts, as it was done in \cite{Popov}
       \beq \label{dv}
       G_{\mbox{\tiny E}}(x^{\mu},\tilde x^{\nu})
       =G^{\mbox{\tiny LFC}}_{\mbox{\tiny E}}(x^{\mu},\tilde x^{\nu})
       + G^{\mbox{\tiny HFC}}_{\mbox{\tiny E}}(x^{\mu},\tilde x^{\nu}),
       \eeq
where
       \beq \label{GLFC}
       G^{\mbox{\tiny LFC}}_{\mbox{\tiny E}}(x^{\mu},\tilde x^{\nu})=\frac1{\pi}
       \int_{0}^{1/\lambda_0} d \omega \cos[\omega (\tau-\tilde
       \tau)] \tilde G_{\mbox{\tiny E}}(\omega; x^{a},\tilde x^{b}),
       \eeq
       \beq \label{GHFC}
       G^{\mbox{\tiny HFC}}_{\mbox{\tiny E}}(x^{\mu},\tilde x^{\nu})=
       \frac1{\pi}\int_{1/\lambda_0}^{\infty} d \omega
       \cos[\omega (\tau-\tilde
       \tau)] \tilde G_{\mbox{\tiny E}}(\omega; x^{a},\tilde x^{b})
       \eeq
if the scalar field is at zero temperature and
       \beq \label{GLFCT}
       G^{\mbox{\tiny LFC}}_{\mbox{\tiny E}}(x^{\mu},\tilde x^{\nu})=
       T \tilde G_{\mbox{\tiny E}}(0; x^{a},\tilde x^{b})
       +2T \sum^{n_0-1}_{n=1} \cos [ \omega_n (\tau-\tilde\tau)]
       \tilde G_{\mbox{\tiny E}}(\omega_n; x^{a},\tilde x^{b}),
       \eeq
       \beq \label{GHFCT}
       G^{\mbox{\tiny HFC}}_{\mbox{\tiny E}}(x^{\mu},\tilde x^{\nu})=
       2T \sum^{\infty}_{n=n_0} \cos [ \omega_n (\tau-\tilde\tau)]
       \tilde G_{\mbox{\tiny E}}(\omega_n; x^{a},\tilde x^{b})
       \eeq
if the field is at temperature $T$. Then the expansion  of
$G^{\mbox{\tiny HFC}}_{\mbox{\tiny E}}(x^{\mu},\tilde x^{\nu})$ in
terms of powers of a small parameter
       \beq \label{wkbpr}
       \varepsilon_{\mbox{\tiny WKB}}=\frac{\lambda_0}{L} \ll 1
       \eeq
can be obtained by analogy with the methods of evaluation of the
DeWitt-Schwinger expansion. If the field is at temperature $T$
then
      \beq
      \lambda_0=\frac1{2\pi T n_0}
      \eeq
and
       \beq \label{wkbprT}
       \varepsilon_{\mbox{\tiny WKB}}=\frac{1}{2\pi T n_0 L} \ll 1.
       \eeq
Below the main points of the Bunch and Parker approach \cite{BP}
for obtaining $G^{\mbox{\tiny HFC}}_{\mbox{\tiny E}} (x^{\mu},
\tilde x^{\nu})$ are outlined.


\section{High-frequency contribution to $\left<
\varphi^2 \right>$ }


Let us introduce Riemann normal coordinates $y^a$ in 3D space with
origin at the point $\tilde x^a$ \cite{Pet}. In these coordinates
one has
        \beq
        g_{ab}(y^a)=\eta_{ab}-\frac13 R_{acbd}\, y^c y^d+ O(y^3),
        \eeq
        \beq
        g^{ab}(y^a)=\eta^{ab}+\frac13 R^{a \ b}_{\ c \ d}
        \, y^c y^d+ O(y^3),
        \eeq
        \beq \label{exg}
        g^{(3)}(y^a)=1-\frac13 R_{ab} y^a y^b + O(y^3),
        \eeq
where the coefficients here and below are evaluated at $y^a=0$
(i.e. at the point $\tilde x^a$), $\eta_{ab}$ denotes the
three-dimensional Euclidean metric. All indices are raised and
lowered with the metric $\eta _{ab}$. Defining $\overline{G}
(\omega;y^a)$ by
        \beq \label{Gw}
        \overline G(\omega;y^a)=\sqrt{g^{(3)}(y)} \tilde
        G_{\mbox{\tiny E}}(\omega;y^a)
        \eeq
and retaining in \Ref{beq} only terms with coefficients involving
two derivatives of the metric or fewer one finds that $\overline G
(\omega;y^a)$ satisfies the equation
        \beq
        \eta^{ab}\frac{\partial^2 \overline G}{\partial y^a \partial
        y^b}-\omega^2 \overline G +\frac13 R_{\ c \ d}^{a \ b} \, y^c y^d
        \frac{\partial^2 \overline G}{\partial y^a \partial y^b} -
         \left[ m^2 +\left(\xi-\frac13 \right)R \right] \overline
         G = -\delta^{(3)}(y).
        \eeq
Note that quantities $R_{abcd}, R_{ab}, R$ evaluated in metric
\Ref{metric} and those, evaluated in 3D metric $g_{ab}$ coincide.
Let us present
        \beq \label{Gy}
        \overline G(\omega;y^a)=\overline G_0(\omega;y^a) + \overline
        G_1(\omega;y^a) + \overline G_2(\omega;y^a)+\dots,
        \eeq
where $\overline G_i(\omega;y^a)$ has a geometrical coefficient
involving $i$ derivatives of the metric at point $y^a=0$. Then
these functions satisfy the equations
        \beq
        \eta^{ab}\frac{\partial^2 \overline G_0(\omega;y^a)}
        {\partial y^a \partial y^b}-\omega^2 \overline
        G_0(\omega;y^a)=-\delta^{(3)}(y),
        \eeq
        \beq
        \eta^{ab}\frac{\partial^2 \overline G_1(\omega;y^a)}
        {\partial y^a \partial y^b}-\omega^2 \overline
        G_1(\omega;y^a)=0,
        \eeq
        \bear \label{G2}
        \eta^{ab}\frac{\partial^2 \overline G_2(\omega;y^a)}
        {\partial y^a \partial y^b}-\omega^2 \overline
        G_2(\omega;y^a)+\frac13 R_{\ c \ d}^{a \ b} \, y^c y^d
        \frac{\partial^2 \overline G_0(\omega;y^a)}{\partial y^a
        \partial y^b} \nn -
         \left[ m^2 +\left(\xi-\frac13 \right)R \right] \overline
         G_0(\omega;y^a)=0.
        \ear
The function $G_0(\omega;y^a)$ satisfies the condition
        \beq
        R_{\ c \ d}^{a \ b} \, y^c y^d \frac{\partial^2
        \overline G_0(\omega;y^a)}{\partial y^a \partial y^b} -
        R^a_b y^b \frac{\partial \overline G_0(\omega;y^a)}
        {\partial y^a}=0,
        \eeq
since $G_0(\omega;y^a)$ is the function only of $\eta_{ab} y^a
y^b$ \cite{BP}. Therefore Eq. \Ref{G2} may be rewritten
        \bear \label{G2n}
        \eta^{ab}\frac{\partial^2 \overline G_2(\omega;y^a)}
        {\partial y^a \partial y^b}-\omega^2 \overline
        G_2(\omega;y^a)+\frac13 R^a_{\ b} \, y^b
        \frac{\partial \overline G_0(\omega;y^a)}{\partial y^a } \nn -
        \left[ m^2 +\left(\xi-\frac13 \right)R \right] \overline
        G_0(\omega;y^a)=0.
        \ear

Let us introduce the local momentum space associated with the
point $y^a=0$ by making the 3-dimensional Fourier transformation
        \beq \label{ms}
        \overline G_i(\omega;y^a)= \int \!\!\!\!\int \limits_{-\infty}
        ^{+\infty}\!\!\!\!\int \frac{dk_1 dk_2 dk_3}{(2 \pi)^3}
        \exp({i k_a y^a}) \overline G_i(\omega;k^a).
        \eeq
It is not difficult to see that
        \beq \label{G0k}
        \overline G_0(\omega;k^a)=\frac{1}{k^2+\omega^2},
        \eeq
        \beq \label{G1k}
        \overline G_1(\omega;k^a)=0,
        \eeq
where $k^2=\eta^{ab}k_a k_b$. In momentum space Eq.\Ref{G2n} gives
        \beq
        -(k^2+\omega^2)\overline G_2(\omega;k^a)-\frac13 R^a_b k_a \frac{\partial
        \overline G_0(\omega;k^a)}{\partial k_b}-\left( m^2 +\xi R
        \right) \overline G_0(\omega;k^a)=0.
        \eeq
Hence
        \beq \label{G2k}
        \overline G_2(\omega;k)=\frac{-m^2-\xi R}{(k^2+\omega^2)^2} +\frac23
        \frac{R^{ab} k_a k_b}{(k^2+\omega^2)^3}.
        \eeq
Substituting \Ref{ms}, \Ref{G0k}, \Ref{G1k}, \Ref{G2k} in \Ref{Gy}
and integrating leads to
        \bear
        \overline G(\omega;y^a)&=&\int \!\!\!\!\int \limits_{-\infty}
        ^{+\infty}\!\!\!\!\int \frac{dk_1 dk_2 dk_3}{(2 \pi)^3}
        \exp({i k_a y^a})\left[\frac1{k^2+\omega^2}
        -\frac{(m^2+\xi R)}{(k^2+\omega^2)^2}
        +\frac{2 \ R^{ab}k_a k_b}{3 (k^2+\omega^2)^3}  \right] \nn
        &=& \frac{1}{8 \pi \exp (|\omega| y)} \left[\frac{-m^2-(\xi
        -1/6)R}{| \omega|} +\frac2y -\frac{R_{ab} y^a y^b}{6 \
        y}\right],
        \ear
where $y=\sqrt{\eta_{ab}y^a y^b}$. Using the definition of
$\overline G(\omega;y^a)$ \Ref{Gw} and expansion \Ref{exg} one
finds
        \bear \label{Gwy}
        \tilde G_{\mbox{\tiny E}}(\omega;y^a)&=&\left(1
        +\frac16 R_{ab}y^ay^b\right)
        \overline G(\omega;y^a)\nn &=& \frac{1}{8 \pi
        \exp (|\omega| y)} \left[\frac{-m^2-(\xi -1/6)R}{|\omega|}
        +\frac2y +\frac{R_{ab} y^a y^b}{6 \ y}\right].
        \ear
The necessary condition for the validity of this approximation is
        \beq
        \omega > \frac1{\lambda_0} \gg \frac1L.
        \eeq
Hence one can evaluate $G^{\mbox{\tiny HFC}}_{\mbox{\tiny E}}$ if
the field is at zero temperature
       \bear \label{hfc}
       G^{\mbox{\tiny HFC}}_{\mbox{\tiny E}}(\triangle \tau,
       y^a)&=&\frac1{\pi}\int_{1/\lambda_0}^{\infty} d \omega
       \cos[\omega \triangle \tau] \tilde
       G_{\mbox{\tiny E}}(\omega; y^a)\nn
       &=&\frac1{8 \pi^2} \left\{ \left[m^2+\left(\xi-\frac16\right)R
       \right] \left[C+\frac12 \ln \left|\, \frac{(y^2 +\triangle
       \tau^2)}{\lambda_0^2}\, \right| \right]\right.\nn
       &&\left. -\frac{2}{y \lambda_0}+\frac1{\lambda_0^2}
       +\frac2{(y^2 +\triangle \tau^2)}
       +\frac{R_{ab}y^a y^b}{6 (y^2 +\triangle \tau^2)}
       \right\},
       \ear
where $C$ is Euler's constant, $\triangle \tau=\tau-\tilde \tau$.

If the field is at temperature $T$ then the necessary condition
for the validity of expression \Ref{Gwy} is
        \beq
        n \geq n_0 \gg \frac1{2 \pi T L}
        \eeq
and
       \bear\label{HFCT}
       G^{\mbox{\tiny HFC}}_{\mbox{\tiny E}}(\triangle \tau,
       y^a)&=& 2T \sum^{\infty}_{n=n_0} \cos
       [\omega_n \triangle \tau] \tilde G_{\mbox{\tiny E}}
       (\omega_n; x^{a},\tilde x^{b})\nn
       &=&\frac1{8 \pi^2} \left\{ \left[m^2+\left(\xi-\frac16\right)R
       \right] \left[C+\frac12 \ln \left|\, (y^2 +\triangle
       \tau^2)(2 \pi T)^2\, \right| +\psi(n_0) \right]\right.\nn
       && -2 \pi T\left(n_0-\frac12 \right)\frac{2}{y}
       +(2 \pi T)^2\left(n_0^2-n_0+\frac16 \right) \nn &&
       \left. +\frac2{(y^2 +\triangle \tau^2)}
       +\frac{R_{ab}y^a y^b}{6 (y^2 +\triangle \tau^2)}
       \right\},
       \ear
where $\psi(z)$ is the logarithmic derivative of the gamma
function (i.e., the digamma function). The Plana sum formula \cite
{WW} has been used to compute the sums over $n$ in the last
expression.


\section{Low-frequency contribution to $\left<
\varphi^2 \right>$ and renormalization procedure}


The behavior of low-frequency modes is determined by the boundary
conditions and the topological structure of the spacetime. As an
example, let us consider the evaluation of the low-frequency
contribution to $\left< \varphi^2 \right>$ on the background of
the small perturbed flat spacetime in a quantum state
corresponding to Minkowski vacuum at the asymptotic. If the
characteristic scale of the gravitational field inhomogeneity
$\lambda$ satisfies the condition
     \beq
     \frac{\lambda }{\lambda_0} \ll 1 \qquad
     \left(\mbox{or } \lambda T n_0 \ll 1 \right),
     \eeq
the low-frequency contributions to $\langle \varphi^2 \rangle$ can
be expanded in terms of powers of this small parameter. Below the
zeroth-order term of this expansion will be used for approximation
of the low-frequency contributions to $\langle \varphi^2 \rangle$.
This means that we choose the long-wave modes approximately
coincident with long-wave modes of Minkowski vacuum. For these
modes in a zero temperature quantum state
      \bear
      G^{\mbox{\tiny LFC}}_{\mbox{\tiny E}}(\triangle \tau,
       y^a)&=&\frac{1}{(2\pi)^4}\int^{1/\lambda_0}_{-1/\lambda_0}
      d \omega \int \!\!\!\!\int \limits_{-\infty}
      ^{\infty}\!\!\!\!\int
       d^3p \frac{\exp\left(i \omega \triangle \tau
      +i p_{\alpha} y^{\alpha}\right)}
      {\left(\omega^2 +p_1^2 +p_2^2 +p_3^2+m^2\right)}
      \nn &=&\frac{1}{4 \pi^3} \int ^{1/\lambda_0}_{-1/\lambda_0}
      d\omega e^{i \omega
      \triangle \tau} \int^{\infty}_{0} d p \frac{p \sin(p \,
      y)}{y \left(\omega^2 +p^2 +m^2
      \right)} \nn &=& \frac{1}{8 \pi^2} \int^{1/\lambda_0}_{-1/\lambda_0}
      d\omega e^{i \omega \triangle \tau} \frac{\exp \left(-y
      \sqrt{\omega^2 +m^2} \right)}{y}.
      \ear
In the limit $\triangle \tau \rightarrow 0,\ y \rightarrow 0$
      \bear
      G^{\mbox{\tiny LFC}}_{\mbox{\tiny E}}(\triangle \tau,
       y^a)&=&\frac{1}{8 \pi^2} \int ^{1/\lambda_0}_{-1/\lambda_0}
      d\omega e^{i \omega \triangle \tau}\left[\frac{1}{y}-
      \sqrt{\omega^2+m^2}+O(y) \right]\nn &=&
      \frac{1}{8 \pi^2} \left[2 \frac{\sin (\triangle \tau/\lambda_0)}
      {y \,\triangle \tau} -\frac{1}{\lambda_0}\sqrt{\frac{1}{\lambda_0^2}+m^2}
      \right. \nn &&\left.-m^2
       \ln \left| \frac{1/\lambda_0+\sqrt{1/\lambda_0^2+m^2}}{m}
      \right| +O(y)\right].
      \ear
If we take into account conditions \Ref{conm} and \Ref{wkbpr},
i.e.
      \beq
      m \ll \frac1{\lambda_0},
      \eeq
then
      \bear \label{lfc}
      G^{\mbox{\tiny LFC}}_{\mbox{\tiny E}}(\triangle \tau, y^a)
      &=& \frac{1}{8 \pi^2 } \left\{\frac{2}{y \lambda_0}
      \left[1-O\left(\frac{\triangle \tau^2}{\lambda_0^2} \right) \right]
      -\frac{1}{\lambda_0^2} \right. \nn && \left.
      -\frac{m^2}{2}\left[1 + \ln \left|
      \frac{4}{\lambda_0^2 m^2}\right| +O\left(\lambda_0^2 m^2
      \right) \right] \right\} +O\left(\frac{\lambda^2}{\lambda_0^4}
      \right).
      \ear
The analogous calculations for the temperature quantum state give
      \bear
      G^{\mbox{\tiny LFC}}_{\mbox{\tiny E}}(\triangle \tau, y^a)
      &=& T \tilde G_{\mbox{\tiny E}}(0; y^{a}))
      +2T \sum^{n_0-1}_{n=1} \cos [ \omega_n \triangle \tau]
       \tilde G_{\mbox{\tiny E}}(\omega_n; y^{a}) \nn
      &=& \frac{1}{8 \pi^2 } \left\{ 2 \pi T
      \left(n_0-\frac12 \right) \frac{2}{y}
      \left[1-O\left(T^2 n_0^2 \triangle \tau^2 \right) \right]
      -(2 \pi T)^2 (n_0^2 -n_0)\right.\nn &&
      \left.-\frac{m^2}{2}\left[1 -\frac1{n_0}
      +\ln \left| \frac{4 (2 \pi T)^2 n_0^2}{m^2}\right|
      +O\left({m^2}{T^2 n_0^2}
      \right) \right] \right\} +O\left(\lambda^2 T^4 n_0^4 \right).
      \ear

The renormalization of $\langle \varphi^2 \rangle_{unren}$ is
achieved by subtracting the renormalization counterterm
\cite{Chris} and then letting $\tilde x^{\mu}\rightarrow x^{\mu} $
      \beq \label{reg}
      \left< \varphi^2 \right>_{ren}=
      \lim_{\tilde x^{\mu} \rightarrow x^{\mu}}
      \left[\left< \varphi^2 \right>_{unren}-
      \left< \varphi^2 \right>_{\mbox{\tiny DS}}\right],
      \eeq
where
      \bear
      \phisq_{\mbox{\tiny DS}} &=& \frac1{8\pi^2\sigma}+\frac1{8\pi^2}
      \left[m^2+\left(\xi-\frac16\right)R\right]
      \left[C+\frac12\ln\left( \frac{\mds^2|\sigma|}{2}
      \right)\right] \nn && -\frac{m^2}{16\pi^2}
      +\frac1{96\pi^2}R_{\mu\nu}
      \frac{\sigma^\mu \sigma^\nu}{\sigma},
      \ear
$\sigma$ is one-half the square of the distance between the points
$x$ and $\tilde x$ along the shortest geodesic connecting them,
$\sigma^{\mu}$ is the covariant derivative of $\sigma$, and the
constant $\mds$ is equal to the mass $m$ of the field for a
massive scalar field. For a massless field $\mds$ is an arbitrary
parameter due to the infrared cutoff in $\left<
\varphi^2\right>_{\mbox{\tiny DS}}$. A particular choice of the
value of $\mds$ corresponds to a finite renormalization of the
coefficients of terms in the gravitational Lagrangian and must be
fixed by experiment or observation. The details of the
calculations of $\phisq_{\mbox{\tiny DS}}$ are discussed in
Appendix:
      \bear \label{ds2}
      \phisq_{\mbox{\tiny DS}} &=& \frac1{8\pi^2} \left\{
      \frac2{y^2 +\triangle \tau^2}+
      \left[m^2+\left(\xi-\frac16\right)R\right]
      \left[C+\frac12\ln\left( \frac{\mds^2}{4}
      \left|\, y^2 +\triangle \tau^2 \right|
      \right)\right]\right. \nn && \left.-\frac{m^2}{2}
      +\frac{R_{ab}y^a y^b}{6(y^2 +\triangle \tau^2)}
      \right\}.
      \ear


\section{Results}


Using Eqs.~\Ref{phi2}, \Ref{dv}, \Ref{hfc}, \Ref{lfc}, \Ref{reg}
and \Ref{ds2} one finds
      \bear \label{T=0}
      \left< \varphi^2 \right>_{ren}
      &=&\lim_{\triangle \tau \rightarrow 0 \atop y \rightarrow 0}
      \left[G^{\mbox{\tiny LFC}}_{\mbox{\tiny E}}(\triangle \tau, y^a)
      +G^{\mbox{\tiny HFC}}_{\mbox{\tiny E}}(\triangle \tau, y^a)
      -\left< \varphi^2 \right>_{\mbox{\tiny DS}}\right] \nn
      &=&\frac{R}{16 \pi^2} \left(\xi-\frac16 \right)
      \ln \left|\frac{4}{\lambda_0^2 m_{\mbox{\tiny DS}}^2}\right|
      +O\left(\frac{\varepsilon^2_{\mbox{\tiny WKB}}}{L^2} \right)
       +O\left(\frac{\lambda^2}{\lambda_0^4} \right).
      \ear
If the field is at temperature $T$ then
      \bear \label{Tn}
      \left< \varphi^2 \right>_{ren}
      &=&\frac{R}{16 \pi^2} \left(\xi-\frac16 \right) \left[
      \ln \left|\frac{4 (2 \pi T)^2}{ m_{\mbox{\tiny DS}}^2}\right|
      +2 \psi (n_0) \right] \nn && +\frac{(2 \pi T)^2}{48 \pi ^2}
      +\frac{m^2}{8 \pi^2} \left[\psi (n_0) -\ln (n_0) +\frac1{2 n_0}
      \right]
      +O\left(\frac{\varepsilon^2_{\mbox{\tiny WKB}}}{L^2} \right)
       +O\left(\lambda^2 T^4 n_0^4\right).
      \ear
Let us cite the conditions of the validity of expressions
\Ref{T=0}, \Ref{Tn} once more
      \beq
      m \lsim \frac{1}{L},
      \eeq
      \beq
      \lambda \ll \lambda_0 \ll L \quad \left( \mbox{or \ }
      \lambda \ll \frac1{2 \pi T n_0} \ll L \right).
      \eeq
If also
      \beq
      n_0 \gg 1,
      \eeq
the digamma function $\psi(n_0)$ is given by
      \beq
      \psi (n_0)= \ln (n_0) -\frac1{2 n_0}
      +O\left(\frac1{n_0^2} \right).
      \eeq
Hence the expression \Ref{Tn} can be rewritten as
      \bear \label{Tn0}
      \left< \varphi^2 \right>_{ren}
      &=&\frac{R}{16 \pi^2} \left(\xi-\frac16 \right)
      \ln \left|\frac{4 (2 \pi T)^2 n_0^2}{ m_{\mbox{\tiny DS}}^2}\right|
      +\frac{(2 \pi T)^2}{48 \pi ^2}
      +O\left(\frac{1}{n_0 L^2} \right) \nn &&
      +O\left(\frac{\varepsilon^2_{\mbox{\tiny WKB}}}{L^2} \right)
       +O\left(\lambda^2 T^4 n_0^4\right).
      \ear

The presence of the arbitrary parameter $\lambda_0$ in the
expressions {\Ref{T=0} is a generic feature of local approximation
schemes \cite{AHS,GAC,BF,PS,Popov,BFNS,FSZ}. For a conformally
invariant field this parameter can be absorbed into the definition
of constant $\mds$.

Note, that the approximation which corresponds to the analytical
approximation of Anderson, Hiscock, and Samuel \cite{AHS} for
$\left< \varphi^2 \right>$ in case of ultrastatic asymptotically
flat spacetime can be obtained by one use of the high-frequency
approximation of $\tilde G_{\mbox{\tiny E}}(\omega; y^a)$ (see Eq.
\Ref{Gwy}) for all the values of $\omega$. However the use of the
high-frequency approximation of $\tilde G_{\mbox{\tiny E}}(\omega;
y^a)$ for $\omega \ll 1/\lambda_0$ does not seem obvious.
Nevertheless, in case of a conformally coupled massless field such
procedure gives good results in the asymptotically flat region.
\cite{AHS,GAC,BF,BFNS}.

\section*{Appendix}

\setcounter{equation}{0}
\renewcommand{\theequation}{A\arabic{equation}}

In this Appendix the expanding of $\phisq_{\mbox{\tiny DS}}$ in
powers of $x^{\alpha} -\tilde  x^{\alpha}$ is described.

Let points $P(x)$ and $\tilde P(\tilde x)$ be connected by the
shortest geodesic $x^{\alpha} =x^{\alpha}(s)$, where $s$ is the
canonical parameter. The functions $x^{\alpha}(s)$ can be expanded
in the Taylor series about point $\tilde P(\tilde x)$
      \beq \label{set}
      x^{\alpha} =\tilde x^{\alpha} +\frac{1}{1!}
      \frac{dx^{\alpha}}{ds} \triangle \!s
      +\frac{1}{2!} \frac{d^2 x^{\alpha}}{ds^2}
      (\triangle s)^2 +\frac{1}{3!} \frac{d^3
      x^{\alpha}}{ds^3} (\triangle s)^3
      +O\left( (\triangle s)^4 \right),
      \eeq
where the coefficients are evaluated at $\tilde P(\tilde x)$.
Using the geodesic equation
      \beq \frac{d^2 x^{\alpha}}{ds^2}
      +\Gamma ^{\alpha }_{\beta \gamma }
      \frac{dx^{\beta}}{ds}  \frac{dx^{\gamma}}{ds}=0,
      \eeq
one finds
      \beq
      \frac{d^{3}x^{\alpha}}{ds^{3}}=
      \left(-\partial_{\gamma}\Gamma^{\alpha}_{\sigma\beta}
      +2\Gamma^{\alpha}_{\sigma\delta}
      \Gamma^{\delta}_{\beta\gamma}\right)
      \frac{dx^{\sigma}}{ds}\frac{dx^{\beta}}{ds}
      \frac{dx^{\gamma}}{ds},
      \eeq
where $\partial_{\gamma}$ denotes the partial derivative with
respect to $x^{\gamma}$.  Hence Eq.~\Ref{set} can be rewritten
      \bear
      x^{\alpha} &=&\tilde x^{\alpha} +\frac{1}{1!}
      u^{\alpha} \triangle \!s
      +\frac{1}{2!} \left( -\Gamma ^{\alpha }_{\beta \gamma }
      u^{\beta} u^{\gamma} \right) (\triangle s)^2 \nn &&
      +\frac{1}{3!} \left(-\partial_{\gamma}\Gamma^{\alpha}_{\sigma\beta}
      +2\Gamma^{\alpha}_{\sigma\delta}
      \Gamma^{\delta}_{\beta\gamma}\right)
      u^{\sigma} u^{\beta} u^{\gamma} (\triangle s)^3
      +O\left( (\triangle s)^4 \right),
      \ear
where $u^{\alpha}=dx^{\alpha}/ds$. This equation can be inverted
      \beq
      u^{\alpha} \triangle s=\epsilon^{\alpha} +\frac{1}{2}
      \Gamma^{\alpha}_{\gamma\beta} \epsilon^{\gamma}
      \epsilon^{\beta} +\left(\frac{1}{6}
      \Gamma^{\alpha}_{\sigma\delta}\Gamma^{\delta}_{\beta\gamma}
      +\frac{1}{6} \partial_{\gamma}\Gamma^{\alpha}_{\sigma\beta}
      \right)\epsilon^{\sigma} \epsilon^{\beta}\epsilon^{\gamma}
      + O(\epsilon^{4}),
      \eeq
where $\epsilon^{\alpha}=x^{\alpha}-\tilde x^{\alpha}$. If we use
determinations $\sigma^{\mu}$ and $\sigma$ \cite{synge}, we can
write
        \bear
        \sigma (x, \tilde x) &=&\frac12 g_{\alpha \beta} u^{\alpha}
        u^{\beta}(\triangle s)^2 = \frac12 g_{\alpha \beta }
        \epsilon^{\alpha } \epsilon^{\beta }
        +\frac12 g_{\alpha \beta }\Gamma^{\alpha }_{\gamma \delta }
        \epsilon^{\beta } \epsilon^{\gamma } \epsilon^{\delta }
        +\frac16 g_{\alpha \beta } \Gamma ^{\alpha }_{\gamma \delta }
        \Gamma ^{\delta }_{\varepsilon \upsilon }\epsilon^{\beta }
         \epsilon^{\gamma } \epsilon^{\varepsilon }
         \epsilon^{\upsilon } \nn
         &&+\frac16 g_{\alpha \beta }\left( \partial_{\varepsilon}
         \Gamma^{\alpha }_{\gamma \delta} \right) \epsilon^{\beta }
         \epsilon^{\gamma } \epsilon^{\delta } \epsilon^{\varepsilon }
         + \frac18 g_{\alpha\beta } \Gamma ^{\alpha }_{\gamma \delta }
         \Gamma ^{\beta}_{\varepsilon \upsilon }  \epsilon^{\gamma }
         \epsilon^{\delta }\epsilon^{\varepsilon } \epsilon^{\upsilon }
         + O\left(\epsilon^5 \right).
         \ear
Hence the resulting expression for $\left< \varphi^2
\right>_{\mbox{\tiny DS}}$ is
         \bear \label{ds3}
         \left< \varphi^2 \right>_{\mbox{\tiny DS}}&=&
         \frac{1}{4 \pi ^2 } \left[ { \frac{1}{g_{\alpha \beta }
         \epsilon^{\alpha } \epsilon ^{\beta}}
         -\frac{g_{\alpha \beta }\Gamma^{\alpha }_{\gamma \delta }
         \epsilon^{\beta } \epsilon^{\gamma } \epsilon^{\delta }}
         {\left(g_{\alpha \beta }\epsilon ^{\alpha }
         \epsilon ^{\beta}\right)^2} -\left({ \frac13
         g_{\alpha \beta} \Gamma ^{\alpha}_{\gamma \delta}
         \Gamma^{\delta}_{\varepsilon \upsilon}
         +\frac13 g_{\alpha \beta} \partial_{\upsilon}
         \Gamma^{\alpha}_{\gamma
         \varepsilon} }\right. }\right. \nn
         & & \left.{ \left.{+\frac14 g_{\alpha \delta} \Gamma
         ^{\alpha}_{\beta \gamma} \Gamma ^{\delta}_{\varepsilon
         \upsilon}}\right) \frac{\epsilon^{\beta } \epsilon^{\gamma }
         \epsilon^{\varepsilon } \epsilon^{\upsilon }}{\left(g_{\alpha
         \beta } \epsilon ^{\alpha } \epsilon ^{\beta}\right)^2}
         +\frac{\left( g_{\alpha \beta } \Gamma ^{\alpha }_{\gamma
         \delta }\epsilon^{\beta }  \epsilon^{\gamma }
         \epsilon^{\delta } \right)^2 } {\left(g_{\alpha \beta }
         \epsilon ^{\alpha } \epsilon^{\beta} \right)^3}
         } \right]  \nonumber \\
         & &+\frac{m^2+\left( \xi - \displaystyle{\frac{1}{6}}
         \right) R}{8 \pi ^2} \left[C+\frac12 \ln \left(
         \frac{ m_{\mbox{\tiny DS}}^2}{4} \left|{g_{\alpha \beta }
         \epsilon ^{\alpha }  \epsilon ^{\beta}} \right|
         \right)  \right] \nonumber \\
         & &- \frac{m^2}{16 \pi ^2} + \frac{1}{48 \pi ^2}
         \frac{R_{\alpha\beta } \epsilon ^{\alpha }
         \epsilon ^{\beta}}{g_{\alpha \beta }
         \epsilon ^{\alpha } \epsilon ^{\beta}}
         + O\left(\epsilon \right).
         \ear
In coordinates $\tau, \ y^a$, where $y^a$ are the Riemann normal
coordinates with origin at the point $\tilde x^a$ in 3D space.
        \beq
        \epsilon^0 =\triangle \tau, \ \epsilon^a=y^a.
        \eeq
At point $\tilde P$ in these coordinates also \cite{Pet}
        \beq
        \Gamma^{\alpha}_{\beta \gamma}=0 \
        \eeq
and
        \beq
        g_{\alpha \beta} \left( \partial_{\upsilon}
        \Gamma^{\alpha}_{\gamma \delta} \right)
        \epsilon^{\beta} \epsilon^{\gamma}
        \epsilon^{\delta} \epsilon^{\upsilon} =
        \frac{\eta_{ab}}{6}\left(R^a_{\ cde} +R^a_{\ dce} \right)
        y^b y^c y^d y^e =0.
        \eeq
After the substitution of these expressions into \Ref{ds3} one
finds that $\left< \varphi^2 \right>_{\mbox{\tiny DS}}$ has the
form \Ref{ds2}.

\section*{Acknowledgments}

I would like to thank S. V. Sushkov and N. R. Khusnutdinov for
helpful discussions. This work was supported by the Russian
Foundation for Basic Research Grant No. 02-02-17177 and by the
SRPED Foundation of Tatarstan Republic Grant No. 06-6.5-110.


\end{document}